\documentclass[a4paper,10pt,twocolumn]{article}
\usepackage[dvips]{graphicx}

\begin{document}

\author{
\\
\\
\\
Maciej Matyka\thanks{Student of Computational Physics subsection
of Theoretical Physics at University of Wroclaw in Poland.
Departament of Physics and Astronomy.}
\\
email: maq@panoramix.ift.uni.wroc.pl\\
\\\\
\\
\\
Exchange Student\\
at\\
University of Linkoping
 }

\title{Incompressible Couette Flow \thanks{Thanks for Grzegorz Juraszek (for English languague checking).}}
\date{}


\maketitle \thispagestyle{empty}

\begin{abstract}

This project work report provides a full solution of simplified
Navier Stokes equations for The Incompressible Couette Problem.
The well known analytical solution to the problem of
incompressible couette is compared with a numerical solution. In
that paper, I will provide a full solution with simple C code
instead of MatLab or Fortran codes, which are known. For discrete
problem formulation, implicit Crank-Nicolson method was used.
Finally, the system of equation (tridiagonal) is solved with both
Thomas and simple Gauss Method. Results of both methods are
compared.

\end{abstract}
\bigskip\bigskip


\section{Introduction}

\begin{figure}[h]
\centering
\includegraphics[scale=1.0]{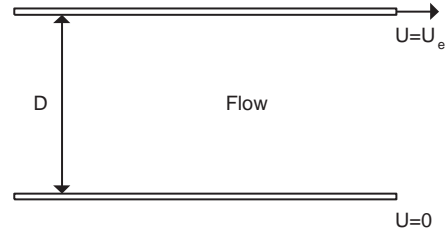}
\caption{Schematic representation of Couette Flow Problem }
\label{schema1}
\end{figure}

Main problem is shown in figure (\ref{schema1}). There is viscous
flow between two parallel plates. Upper plate is moving in x
direction with constans velocity $(U=U_e)$. Lower one is not
moving $(U=0)$. We are looking for a solution to describe
velocity vector field in the model (between two plates).

\section{Fundamental Equations}

Most of incompressible fluid mechanics (dynamics) problems are
described by simple Navier-Stokes equation for incompressible
fluid velocity, which can be written with a form:

\begin{equation} \label{n-s equation}
\frac{\partial \vec {u}}{\partial t} = ( - \vec {u}\vec {\nabla
})\vec {u} - \vec {\nabla }\varphi + \upsilon \vec {\nabla
}^2\vec {u} + \vec {g},
\end{equation}

where $\varphi$ is defined is defined as the relation of pressure
to   density:

\begin{equation}
\varphi = \frac{p}{\rho}
\end{equation}

and $\upsilon$ is a kinematics viscosity of the fluid.

We will also use a continuity equation, which can be written as
follows \footnote{I assume constans density of the fluid.}:

\begin{equation} \label{continuity equation}
D=\vec{\nabla}\cdot \vec{v}=0
\end{equation}

Of course, in a case of couette incompressible flow we will use
several simplifications of (\ref{n-s equation}).

\section{Mathematical Formulation of the Couette Problem}

Incompressible couette problem is not needed to solve full
Navier-Stokes equations. There is no external force, so first
simplification of (\ref{n-s equation}) will be:

\begin{equation} \label{n-s equation1}
\frac{\partial \vec {u}}{\partial t} = ( - \vec {u}\vec {\nabla
})\vec {u} - \vec {\nabla }\varphi + \upsilon \vec {\nabla
}^2\vec {u},
\end{equation}

In \cite{ander} there can be found easy proof that in couette
problem there are no pressure gradients, which means that:

\begin{equation}
\vec {\nabla }\varphi = 0,
\end{equation}

We will ignore a convection effects so, equation (\ref{n-s
equation1}) can be written with a form:

\begin{equation} \label{n-s equation2}
\frac{\partial \vec {u}}{\partial t} =  \upsilon \vec {\nabla
}^2\vec {u}
\end{equation}

Now we have simple differential equation for velocity vector
field. That equation is a vector type and can be simplified even
more. Let us write continuity equation (\ref{continuity
equation}) in differential form. Let $\vec{u}=(u,v)$, then
continuity equation can be expanded as follows\footnote{Only with
assumption of non compressible fluid.}:

\begin{equation}
\frac{\partial u}{\partial x}+\frac{\partial v}{\partial y}=0
\end{equation}

We know that there is no $u$ velocity component gradient along x
axis (symmetry of the problem), so:

\begin{equation}
\frac{\partial u}{\partial x}=0
\end{equation}

Evaluation of Taylor series\footnote{Those simple expressions can
be found in \cite{ander}, chapter 9.2.} at points $y=0$ and $y=D$
gives us a proof that only one possible and physically correct
value for y component of velocity $\vec{u}$ is:

\begin{equation} \label{yvelocity}
v=0
\end{equation}

Because of (\ref{yvelocity}) equation (\ref{n-s equation2}) can
be written as follows:

\begin{equation} \label{lastequation}
\frac{\partial u}{\partial t}=\upsilon \frac{\partial^2
u}{\partial y^2}
\end{equation}

Our problem is now simplified to mathematical problem of solving
equations like (\ref{lastequation}). That is now a governing
problem of incompressible couette flow analysis.

\subsection{Analytical soulution}

Analytical solution for velocity profile of steady flow, without
time-changes (steady state) can be found very easily in an
equation:

\begin{equation} \label{lastequation - steady state1}
\upsilon \frac{\partial^2 u}{\partial y^2}=0
\end{equation}

And without any changes of viscosity $\upsilon$ it can be written
in form:

\begin{equation} \label{lastequation - steady state2}
\frac{\partial^2 u}{\partial y^2}=0
\end{equation}

After simple two times integration of equation (\ref{lastequation
- steady state2}) we have analytical solution function of
(\ref{lastequation - steady state2}):

\begin{equation}\label{analitsolut}
u=c_1\cdot y+c_2
\end{equation}

where $c_1$ and $c_2$ are integration constans.

\subsection{Boundary Conditions for The Analytical Solution}

Simple boundary conditions are provided in that problem. We know
that:

\begin{equation}
u = \left\{
\begin{array}{ll}
0 & \textrm{y=0}\\
u_e & \textrm{y=D}
\end{array} \right.
\end{equation}

Simple applying it to our solution (\ref{analitsolut}) gives a
more specified one, where $c_1=\frac{u_e}{D}$ and $c_2=0$:

\begin{equation}\label{analitsolut2}
u=\frac{u_e}{D} \cdot y
\end{equation}

It means, that a relationship between $u$ and $y$ is linear. A
Better idea to write that with mathematical expression is:

\begin{equation}\label{analitsolut3}
\frac{u}{y}=\frac{u_e}{D}
\end{equation}

Where $\frac{u_e}{D}$ is a constans for the problem (initial $u$
velocity vs size of the model).

\section{Numerical Solution}
\subsection{Non-dimensional Form}

Let us define some new non-dimensional variables\footnote{Exacly
the same, like in \cite{ander}.}:

\begin{equation} \label{dimensionless}
u = \left\{
\begin{array}{lll}
u^{'} & = & \frac{u}{u_e} \\
y^{'} & = & \frac{y}{D} \\
t^{'} & = & \frac{t}{D/u_e}
\end{array}
\right.
\end{equation}

Now let us place these variables into equation
(\ref{lastequation}) and we have now a non-dimensional equation
written as follows:

\begin{equation}
\rho \frac{\partial u/u_e}{(t\cdot u_e)/D}\left(
\frac{u^2_e}{D}\right)=\upsilon \frac{\partial^2(u/u_e)}{\partial
(y/D)^2}\left(\frac{u_e}{D^2}\right)
\end{equation}

Now we replace all the variables to nondimensional, like defined
in (\ref{dimensionless}):

\begin{equation}
\rho \frac{\partial u^{'}}{\partial t^{'}}\left(
\frac{u^2_e}{D}\right)=\upsilon \frac{\partial^2 u^{'}}{\partial
y^{'2}}\left(\frac{u_e}{D^2}\right)
\end{equation}

Now we will remove all $^{'}$ chars from that equation (only for
simplification of notation), and the equation
becomes\footnote{Constans simplification also implemented} to:

\begin{equation}
\frac{\partial u}{\partial t}=\left( \frac{\upsilon}{D \rho u_e}
\right)\frac{\partial^2 u}{\partial y^{2}}
\end{equation}

In that equation Reynold's number $Re$ appears, and is defined as:

\begin{equation}\label{reynolds}
Re=\frac{D \rho u_e}{\upsilon}
\end{equation}

Where $Re$ is Reynold's number that depends on $D$ height of
couette model. Finally, the last form of the equation for the
couette problem can be written as follows:

\begin{equation}\label{lastnumeq}
\frac{\partial u}{\partial t}=\left( \frac{1}{Re}
\right)\frac{\partial^2 u}{\partial y^{2}}
\end{equation}

We will try to formulate numerical solution of the equation
(\ref{lastnumeq}).

\subsection{Finite - Difference Representation}

In our solution of equation (\ref{lastnumeq}) we will use
Crank-Nicolson technique, so discrete representation of that
equation can be written \footnote{That representation is based on
central discrete differential operators.} as:

\begin{equation} \label{cranck}
u_j^{n+1}=u_j^n+\frac{\Delta t}{2(\Delta y)^2 Re}
(u_{j+1}^{n+1}+u_{j+1}^{n}-2u_{j}^{n+1}+u_{j-1}^{n+1}+u_{j-1}^{n})
\end{equation}

Simple grouping of all terms which are placed in time step (n+1)
on the left side and rest of them - on right side, gives us an
equation which can be written as:

\begin{equation} \label{systemofequation1}
Au^{n+1}_{j-1}+Bu^{n+1}_{j}+Au^{n+1}_{j+1}=K_j
\end{equation}

Where $K_j$ is known and depends only on values $u$ at $n$ time
step:

\begin{equation}
K_j=\left( 1-\frac{\Delta t}{(\Delta y)^2 Re}\right)
u^n_j+\frac{\Delta t}{2(\Delta y)^2 Re}(u^n_{j+1}+u^n_{j-1})
\end{equation}

Constans $A$ and $B$ are defined as follows\footnote{Directly from
equation (\ref{cranck}).}:

\begin{equation}
A=-\frac{\Delta t}{2 (\Delta y)^2 Re}
\end{equation}

\begin{equation}
B=1+\frac{\Delta t}{(\Delta y)^2 Re}
\end{equation}

\subsection{Crank - Nicolson Implicit scheme}

For numerical solution we will use one-dimensional grid points
$(1,\ldots,N+1)$ where we will keep calculated $u$ velocities.
That means $u$ has values from the range
$(u_1,u_2,\ldots,u_{N+1})$. We know (from fixed boundary
conditions), that: $u_1=0$ and $u_{N+1}=0$. Simple analysis of
the equation (\ref{systemofequation1}) gives us a system of
equations, which can be described by matrix equation:

\begin{equation} \label{govequation}
\mathbf{A}\cdot \vec{X}=\vec{Y}
\end{equation}

Where $\mathbf{A}$ is tridiagonal $[N-1]\cdot [N-1]$ matrix of
constant $A$ and $B$ values:

\begin{equation}
\mathbf{A} = \left[
\begin{array}{llllll}
B & A &   &   &   &   \\
A & B & A &   &   &   \\
  &  & \ldots &  &   &   \\
  &  &  & \ldots &  &   \\
  &   &   & A & B & A \\
  &   &   &   & A & B \\
\end{array} \right]
\end{equation}

$\vec{X}$ vector is a vector of $u$ values:

\begin{equation}
\vec{X}=[u_1,u_2,\ldots,u_{N+1}]
\end{equation}

$\vec{Y}$ vector is a vector of constans $K_j$ values:

\begin{equation}
\vec{Y}=[K_1,\ldots,K_{N+1}]
\end{equation}

\section{Solving The System of Linear Equations}

Now the problem comes to solving the matrix - vector equation
(\ref{govequation}). There are a lot of numerical methods for
that\footnote{Especially for tridiagonal matrices like
$\mathbf{A}$}, and we will try to choose two of them: Thomas and
Gauss method. Both are very similar, and I will start with a
description of my implementation with the simple Gauss method.

\subsection{Gauss Method}

Choice of the Gauss method for solving system of linear equations
is the easiest way. This simple algorithm is well known, and we
can do it very easily by hand on the paper. However, for big
matrices (big $N$ value) a computer program will provide us with a
fast and precise solution. A very important thing is that time
spent on writing (or implementing, if Gauss procedure was written
before) is very short, because of its simplicity.

I used a Gauss procedure with partial choice of a/the general
element. That is a well known technique for taking the first
element from a column of a matrix for better numerical accuracy.

The whole Gauss procedure of solving a system of equations
contains three steps. First, we are look- ing for the general
element.

After that, when a general element is in the first row (we make an
exchange of rows\footnote{We made it for $\mathbf{A}$ matrix, and
for $\vec{X},\vec{Y}$ too.}) we make some simple calculations
(for every value in every row and column of the matrix)  for the
simplified matrix to be diagonal (instead of a tridiagonal one
which we have at the beginning). That is all, because after
diagonalization I implement a simple procedure (from the end row
to the start row of the matrix) which calculates the whole vector
$\vec{X}$. There are my values of $u_i$ velocity in all the
model\footnote{More detailed description of Gauss method can be
found in a lot of books on numerical methods, like
\cite{potter}.}.

\subsection{Thomas Method}

Thomas' method, described in \cite{ander} is simplified version
of Gauss method, created especially for tridiagonal matrices.
There is one disadvantage of Gauss method which disappears when
Thomas' method is implemented. Gauss method is rather slow, and
lot of computational time is lost, because of special type of
matrix. Tridiagonal matrices contain a lot of free (zero) values.
In the Gauss method these values joins the calculation, what is
useless.

Thomas' simplification for tridiagonal matrices is to get only
values from non-zero tridiagonal part of matrix. Simply applying
a Thomas' equations for our governing matrix equation
(\ref{govequation}) gives us:

\begin{equation}\label{thomas1}
  d_i^{'}=B-\frac{A^2}{d_{i-1}^{'}}
\end{equation}

\begin{equation}\label{thomas1}
  u_i^{'}=u_i-\frac{u_i^{'}A}{d_{i-1}^{'}}
\end{equation}

We know that exact value of $u_M$ is defined as follows:

\begin{equation}
u_M=\frac{u^{'}_M}{B}
\end{equation}

Now solution of the system of equations will be rather easy. We
will use recursion like that:

\begin{equation}
 u_{i-1}=\frac{u^{'}_{i-1}-A\cdot u_i}{B}
\end{equation}

That easy recursion provides us a solution for the linear system
of equations.

\section{Results}

Main results are provided as plots of the function:

\begin{equation}\label{functionres}
  \frac{y}{D}=f\left( \frac{u}{u_e} \right)
\end{equation}

\begin{figure}[h]
\centering
\includegraphics[scale=0.65]{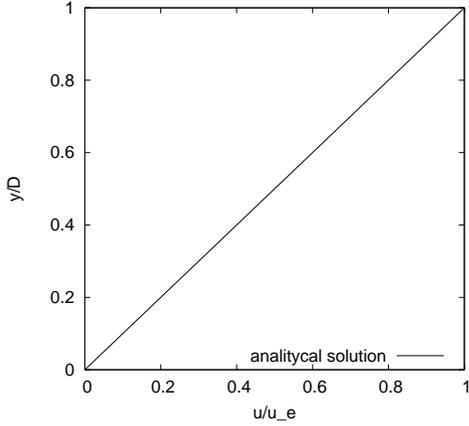}
\caption{Analytical exact solution.} \label{plot1anal}
\end{figure}

In figure (\ref{plot1anal}) there is drawn an analytical solution
to the problem of couette flow. That is linear function, and we
expect that after a several time steps of numerical procedure we
will have the same configuration of velocity field.

\subsection{Different Time Steps}

\begin{figure}[h]
\centering
\includegraphics[scale=0.65]{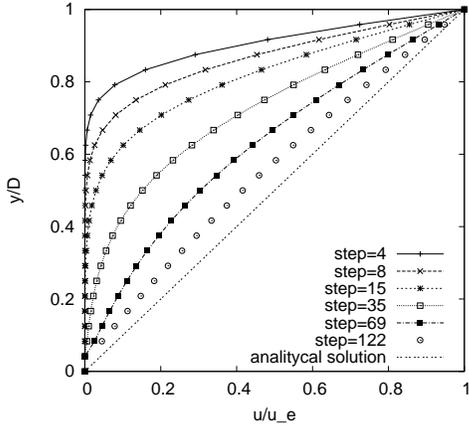}
\caption{Results for different time steps of numerical
calculations.} \label{plot1}
\end{figure}

In figure (\ref{plot1}) there are results of velocity $u$
calculation for several different time steps. Analytical solution
is also drawn there.

As we can see in the figure (\ref{plot1}) - the solution is going
to be same as analytical one. Beginning state (known from boundary
conditions) is changing and relaxing.

\subsection{Results for Different Reynolds Numbers}

\begin{figure}[h]
\centering
\includegraphics[scale=0.65]{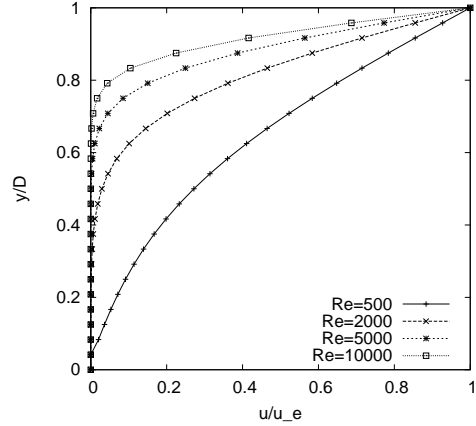}
\caption{Calculation for same time ($t=2500$) and different
Reynold's numbers.} \label{plotreynolds}
\end{figure}

In the figure (\ref{plotreynolds}) there is plot of numerical
calculations for different Reynold's numbers. For example
Reynold's number depends on i.e. viscosity of the fluid, size of
couette model. As it is shown on the plot there is strong
relationship between the speed of the velocity field changes and
Reynold's number. In a couple of words: when Reynolds number
increases - frequency of changes also increases.

\subsection{Results for Different Grid Density}

\begin{figure}[h]
\centering
\includegraphics[scale=0.65]{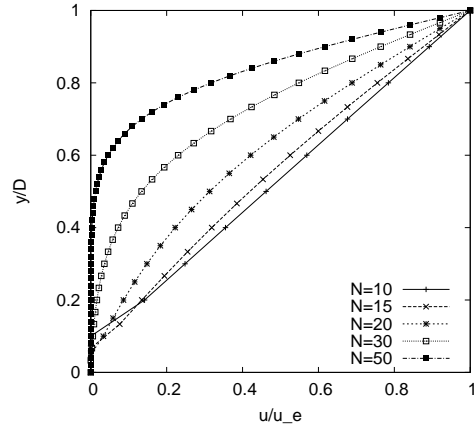}
\caption{Calculation for same time ($t=2500$), same Reynold's
numbers (Re=5000) and different grid density ($N$ number).}
\label{plotgrid}
\end{figure}

In figure (\ref{plotgrid}) there is an example of calculations of
velocity field for different grid density ($N$ number). We see
that there is also strong correlation between grid density, and
speed of changes on the grid. Also, very interesting case $N=10$
shows, that for low density of the grid changes are very fast,
and not accurate.

\subsection{Conclusion}

Solving of Incompressible Couette Problem can be good way to
check numerical method, because of existing analytical solution.
In that report there were presented two methods of solving system
of equations: Gauss and Thomas' method. System of equations was
taken from Crank-Nicolson Implicit scheme. Well known linear
relationships were observed.

\onecolumn

\section{APPENDIX A}
\begin{verbatim}

#include <stdlib.h>
#include <stdio.h>
#include <math.h>

#define N (40)
#define NN (N+1)

void Zamien(double *a, double *b) {
    double c;
    c=*a;   *a=*b;  *b=c;
}

void WypiszMacierz(double A[NN][NN], int n) {
    int i,j;

    for(j=1;j<n;j++)
    {
        for(i=1;i<n;i++)            // show matrix
        {
            printf("%2.4f   ",A[i][j]);
        }
        printf("\n");
    }
}

void Gauss(double A[NN][NN], double *b, double *x, int n) {
    int i,j,k;
    double m;

// Gauss Elimination

    for(i=0;i<n;i++)
    {
        // Step #1: Change governing element

        m=fabs(A[i][i]);
        k=i;

        for(j=i+1;j<n;j++)
        if(fabs(A[i][j])>m)
        {
            m=fabs(A[i][j]);
            k=j;
        }
        if(k!=i)
        for(j=0;j<n;j++)
        {
            Zamien(&A[j][i],&A[j][k]);
            Zamien(&b[i+1],&b[k+1]);
        }

        // Step #2: make it triangle
        for(j=i+1;j<n;j++)
        {
            m = A[i][j]/A[i][i];

            for(k=i;k<n;k++)
                A[k][j] = A[k][j] - m*A[k][i];

            b[j+1] = b[j+1] - m*b[i+1];
        }
    }
// Step#3: Solve now
    for(i=n-1;i>=1;i--)
    {
        for(j=i+1;j<n;j++)
            b[i+1] = b[i+1]-A[j][i]*x[j+1];
        x[i+1] = b[i+1]/A[i][i];
    }
}

int main(void)
{
    double U[N*2+2]={0},A[N*2+2]={0},B[N*2+2]={0},C[N*2+2]={0},D[N*2+2]={0},Y[N*2+2]={0};
// initialization
    double OneOverN = 1.0/(double)N;
    double Re=5000;                     // Reynolds number
    double EE=1.0;                      // dt parameter
    double t=0;
    double dt=EE*Re*OneOverN*2;         // delta time
    double AA=-0.5*EE;
    double BB=1.0+EE;
    int KKEND=1122;
    int KKMOD=1;
    int KK;                             // for a loop
    int i,j,k;                          // for loops too
    int M;                              // temporary needed variable
    double GMatrix[NN][NN]={0};             // for Gauss Elimination
    double test;
    Y[1]=0; // init

// apply boundary conditions for Couette Problem

    U[1]=0.0;
    U[NN]=1.0;

// initial conditions (zero as values of vertical velocity inside of the couette model)

    for(j=2;j<=N;j++)
        U[j]=0.0;

    A[1]=B[1]=C[1]=D[1]=1.0;

    for(KK=1;KK<=KKEND;KK++)
    {
        for(j=2;j<=N;j++)
        {
            Y[j]=Y[j-1]+OneOverN;
            A[j]=AA;
            if(j==N)
                A[j]=0.0;
            D[j]=BB;
            B[j]=AA;
            if(j==2)
                B[j]=0.0;
            C[j]=(1.0-EE)*U[j]+0.5*EE*(U[j+1]+U[j-1]);
            if(j==N)
                C[j]=C[j]-AA*U[NN];
        }

// Gauss
// C[]       -   free
// A[]B[]D[] -   for matrix calculation
// U[]       -   X

// calculate matrix for Gauss Elimination

        GMatrix[0][0]=D[1];
        GMatrix[1][0]=A[1];

        for(i=1;i<N-1;i++)
        {
            GMatrix[i-1][i]=B[i+1];     // GMatrix[1][2]=B[2]
            GMatrix[i][i]=D[i+1];       // GMatrix[2][2]=D[2]
            GMatrix[i+1][i]=A[i+1];     // GMatrix[3][2]=A[2]
        }

        GMatrix[N-2][N-1]=B[N];
        GMatrix[N-1][N-1]=D[N];

        Gauss(GMatrix,C,U,N);           // Gauss solving function

        Y[1]=0.0;
        Y[NN]=Y[N]+OneOverN;

        t=t+dt;                         // time increment
        test=KK % KKMOD;

       if(test < 0.01)                  // print the results
        {
            printf("KK,TIME\n");        // info 1
            printf("%d,%f\n",KK,t);

            printf(",J,Y[J],U[j]\n");   // info 2
            for(j=1;j<=NN;j++)
                printf("%d , %f, %f\n",j,U[j],Y[j]);
                printf("\n \n \n \n");      // for nice view of several datas
        }
     }

    return (1);
}
\end{verbatim}

\section{APPENDIX B}
\begin{verbatim}

#include <stdlib.h>
#include <stdio.h>

#define N (50)
#define NN (N+1)

int main(void)
{
    double U[N*2+2],A[N*2+2],B[N*2+2],C[N*2+2],D[N*2+2],Y[N*2+2];
// initialization

    double OneOverN = 1.0/(double)N;
    double Re=7000;                     // Reynolds number
    double EE=1.0;                      // dt parameter
    double t=0;
    double dt=EE*Re*OneOverN*2;         // delta time
    double AA=-0.5*EE;
    double BB=1.0+EE;
    int KKEND=122;
    int KKMOD=1;
    int KK;                             // for a loop
    int j,k;                            // for loops too
    int M;                              // temporary needed variable

    double test;
    Y[1]=0; // init

// apply boundary conditions for Couette Problem
    U[1]=0.0;
    U[NN]=1.0;

// initial conditions (zero as values of vertical velocity inside of the couette model)
    for(j=2;j<=N;j++)
        U[j]=0.0;
    A[1]=B[1]=C[1]=D[1]=1.0;

    printf("dt=%f,    Re=%f,   N=%d \n",dt,Re, N);

    for(KK=1;KK<=KKEND;KK++)
    {
        for(j=2;j<=N;j++)
        {
            Y[j]=Y[j-1]+OneOverN;
            A[j]=AA;
            if(j==N)
                A[j]=0.0;

            D[j]=BB;
            B[j]=AA;

            if(j==2)
                B[j]=0.0;
            C[j]=(1.0-EE)*U[j]+0.5*EE*(U[j+1]+U[j-1]);

            if(j==N)
                C[j]=C[j]-AA*U[NN];
        }

        // upper bidiagonal form
        for(j=3;j<=N;j++)
        {
            D[j]=D[j]-B[j]*A[j-1]/D[j-1];
            C[j]=C[j]-C[j-1]*B[j]/D[j-1];
        }

        // calculation of U[j]
        for(k=2;k<N;k++)
        {
            M=N-(k-2);
            U[M]=(C[M]-A[M]*U[M+1])/D[M];   // Appendix A
        }

        Y[1]=0.0;
        Y[NN]=Y[N]+OneOverN;
        t=t+dt;                             // time increment
        test=KK % KKMOD;

        if(test < 0.01)                     // print the results
        {
            printf("KK,TIME\n");            // info 1
            printf("%d,%f\n",KK,t);

            printf(",J,Y[J],U[j]\n");       // info 2
            for(j=1;j<=NN;j++)
                printf("%d , %f, %f\n",j,U[j],Y[j]);
                printf("\n \n \n \n");      // for nice view of several datas
        }
    }
    return (1);
}
\end{verbatim}

\end{document}